\documentclass[]{raa-rad}                  
\usepackage{graphicx,times,wrapfig}             
\input{epsf.sty}                        
\input{psfig.sty}                       
\usepackage{color}
\begin{document}

   \title{On the Radiation Problem \textit{of} High Mass Stars
\footnotetext{$^{*}$Supported by the Republic \textit{of} South Africa's  National Research Foundation and the North West University, and Germany's DAAD 
Programme \textit{via} the University \textit{of} K$\ddot{\rm{o}}$ln.}
}
  
   \volnopage{Vol.0 (200x) No.0, 000--000}      
   \setcounter{page}{1}          

   \author{G. G. Nyambuya}
   

\institute{North West University (Potchefstroom Campus), School \textit{of} Physics (Unit \textit{for} Space Research), Private Bag $X6001$, Potchefstroom $2531$, Republic \textit{of} South Africa.; {\it gadzirai@gmail.com}}

\date{Received $4$ Jan. $2010$ ; Accepted $2$ March $2010$}

\abstract{A massive star is defined to be one with {mass} greater than $\sim8-10\mathcal{M}_{\odot}$. Central to the on-going debate on how these objects [massive stars] come into being is the so-called Radiation Problem. For nearly forty {years}, it has been argued that  the radiation field emanating from massive stars  is high enough to cause a  global reversal of direct radial in-fall of  material onto the nascent star. We argue that only in the case of a non-spinning isolated star does the gravitational field of the nascent star overcome the radiation field. An isolated non-spinning star is a non-spinning star without any circumstellar material around it, {and} the gravitational field beyond its surface is described exactly by Newton's inverse square law. The {supposed} fact that massive stars have {a} gravitational field {that is} much stronger than their radiation field is drawn from the analysis  of an isolated massive star. {In} this case the gravitational field is much stronger than the radiation field. This conclusion {has been} erroneously extended to the case of massive stars enshrouded in gas \textit{\&} dust. We find that, for the case of a non-spinning gravitating body where we take into consideration the circumstellar material, that at $\sim8-10\mathcal{M}_{\odot}${, t}he radiation field will not reverse the radial in-fall of matter{,} but {rather} a stalemate between the radiation and gravitational field will be achieved{,  \textit{i.e.}} in-fall is halted but not reversed. This picture is very different from the common picture that is projected and accepted in the popular literature that at $\sim8-10\mathcal{M}_{\odot}$, all the circumstellar material{,} from the surface of the {star right} up to the edge of the {molecular} core, is expected to be swept away by the radiation field. We argue that massive stars should be able to {start their normal stellar processes} if the molecular core from which they form {has} some rotation, because a rotating core exhibits an {Azimuthally Symmetric Gravitational Field} which causes there to be {an accretion} disk and along this  disk{. T}he radiation field {cannot} be much stronger than the gravitational field{,} hence this equatorial accretion disk becomes the channel \textit{via} which the nascent massive star accretes all of its material.
\\
\textbf{PACS (2010):} $97.10.$Bt, $97.10.$Gz, $97.10.$Fy
}

   \authorrunning{G. G. Nyambuya}            
   \titlerunning{ The Radiation Problem \textit{of} Massive Stars Revisited}  

   \maketitle

%

{\renewcommand{\theequation}{\textcolor{blue}{\arabic{equation}}}
 \renewcommand{\thefigure}{(\textcolor{red}{\arabic{figure}})}
 \renewcommand{\thetable}{(\textcolor{red}{\Roman{table}})}
 
%

\section[\textbf{Introduction}]{Introduction}

According {to current} and prevailing wisdom, it is bona-fide scientific knowledge that our current understanding of massive star formation is lacking{. T}his is due to the existing theoretical and observational dichotomy. In the gestation period of a star's life, its mass will grow \textit{via} the in-falling envelope (\textit{i.e.}, circumstellar material) and also through the form{ation of an} accretion disk {lying} along {the plane of it's} equator. As far as our theoretical understanding is concerned, this works well for stars less than about $8-10\mathcal{M}_{\odot}$. In the literature, it is said that the problem of massive stars ($\mathcal{M}_{star}>8-10\mathcal{M}_{\odot}$) arises because as the central {protostar's} mass grows, so does the radiation pressure from it, and at about $8-10\, \mathcal{M}_{\odot}$, the star's radiation pressure becomes powerful enough to halt any further in-fall of matter onto the protostar (Larson \textit{\&} Starfield \cite{larson71}; Kahn \cite{kahn74}; Yorke \textit{\&} Kr$\ddot{\textrm{u}}$gel \cite{yorke74};  Wolfire \textit{\&} Cassinelli \cite{wolfire87};  Palla \textit{\&} Stahler \cite{palla93}; Yorke \cite{yorke02}; Yorke \textit{\&} Sonnhalter \cite{yorke-s03}). So the problem is{:} how does the star continue to accumulate more mass beyond the $8-10\, \mathcal{M}_{\odot}$ limit? If the radiation field really did reverse any further in-fall of matter and protostars exclusively accumulated mass \textit{via} direct radial in-fall of matter onto the nascent star and also \textit{via} the accretion disk, this would set a mass upper limit of $8-10\, \mathcal{M}_{\odot}$ for any star in the Universe. Unfortunately (or maybe fortunately) this is not what we observe. It therefore means that some process(es) responsible for the formation of stars beyond the $8-10\, \mathcal{M}_{\odot}$ limit  must be at work{. A} solution to the problem must be sought because observations dictate {that it} exists. 

If this is the case, \textit{i.e.} the radiation problem really did exist as stated above, and our physics {where} complete \textit{viz} gravitation and radiation transport, {then, the} solution to the conundrum would be to seek a star formation model that overcomes the radiation pressure problem {while} at the sametime allowing for the star to form (accumulate all of its mass) before it exhausts its nuclear fuel. Two such (competing) models have been set-forth{:} ($1$) the Accelerated Accretion Model (AAM) (Yorke \cite{yorke02}, \cite{yorke03}) and, ($2$) the Coalescence Model (CM) (Bonnell et al. \cite{bonnell98}, \cite{bonnell01}, \cite{bonnell04}, \cite{bonnell06}, \cite{bonnell07}; Bonnell \& Bate \cite{bonnell02}). 

The latter scenario, {the CM,} is born out of the observational fact that massive stars are generally found in the {centers} of dense clusters (see \textit{e.g.}  Hillenbrand \cite{hillenbrand97}; Clarke et al. \cite{clarke00}). In these dense environments, the probability of collision of proto-stellar objects is significant, {leading to} the CM.  This model easily by-passes the radiation pressure problem and{,} despite the fact that not a single observation to date has confirmed it (directly or indirectly), it [CM] appears\footnote{This {relies} on the assumption that our understanding of gravitation and radiation transport is complete.} to be the most natural mechanism by which massive stars form given the said observational fact about massive stars and their preferential environment.

The AAM is just a scaled up version of the accepted accretion paradigm applicable to Low Mass Stars (LMSs). This accretion takes place \textit{via} the accretion disk and{,} for the reason mentioned above that the accretion mechanism must be such that it allows for the star to form before it exhausts its nuclear fuel, the accretion {cannot} take place at the same steady rate as in the case of LMSs ($\mathcal{M}\leq \, 3\mathcal{M}_{\odot}$) but must be accelerated and significantly  {higher}. While there exists many examples of massive stars surrounded by accretion disks, one of the chief obstacles  in verifying this paradigm is that examples of HMSs tend to be relatively distant ($>1\, \textrm{kpc}$), deeply embedded, and confused with other emission sources (see \textit{e.g.} Mathews et al. \cite{mathews07}). Additionally, HMSs evolve rapidly, and by the time an unobstructed view of the young star emerges, the disk and outflow structures may have been destroyed. {Consequently}, observations to date have been unable to probe the $10-100\, \textrm{AU}$ spatial scales over which outflows from the accretion disks are expected to be launched and collimated (\textit{e.g.} Mathews et al. \cite{mathews07}).

The other alternative, which is less pursued, would be to seek a physical mechanism that overcomes the radiation pressure problem as has been conducted by the authors Krumholz et al. (\cite{krumholz09}, \cite{krumholz05}). These authors (Krumholz et al. \cite{krumholz09}, \cite{krumholz05}) believe that the radiation problem does not exist because radiation-driven bubbles that block accreting gas are subject to Rayleigh-Taylor instability which occurs anytime a dense, heavy fluid is being accelerated by light{er} fluid{,} for example{,} when a cloud receives a shock, or when a fluid of a certain density floats above a fluid of lesser density, such as dense oil floating on water. The Rayleigh-Taylor instabilities {allow} fingers of dense gas to break into the evacuated bubbles and reach the stellar surface while in addition, outflows from massive stars create optically thin cavities in the accreting envelope. These channel radiation away from the bulk of the gas and reduce the radiation pressure it experiences. In this case, the radiation pressure feedback is not the dominant factor in setting the final size of massive stars and accretion will proceed{,} albeit at much higher rates. Amongst others, the model by the authors Krumholz et al. (\cite{krumholz09}, \cite{krumholz05}) is {ad hoc} rather than natural, in that \textit{Nature} has to make a special arrangement or must configure herself in such a way that massive stars have a way {of starting their normal stellar processes}. Does there not exist a smooth and natural way to {bring massive} stars into {existence}?

In this reading,  we redefine the radiation problem (for the spherically symmetric case) and we do this \textit{via} a subtle and overlooked assumption made in the analysis leading to the radiation problem{:} that the surroundings of the protostar {is} a vacuum (see \textit{e.g.} Yorke \cite{yorke02}; Yorke \textit{\&} Sonnhalter \cite{yorke-s03}; Zinnecker \textit{\&} Yorke \cite{zinnecker07}); surely, this is clearly not true. The researchers Yorke \cite{yorke02}; Yorke \textit{\&} Sonnhalter {\cite{yorke02}};  Zinnecker \textit{\&} Yorke \cite{zinnecker07}; {among} others, hold the view {that from} a theoretical stand-point, the radiation field is stronger than the gravitational field for massive stars hence the in-fall process of material must be reversed; but this conclusion has been reached{,} as will be shown in the next section; after comparing the gravitational field strength at point $r$ of a star in empty space to its radiation field strength at point $r$. In practice, stars are found embedded inside a significant mass of gas and dust. The radiation problem is arguably the most important problem of all in the study of the formation of stars, {thus}, it is important to make sure that this problem is clearly defined and understood.

Having taken into consideration the circumstellar material, we find that at $\sim8-10\mathcal{M}_{\odot}$, the radiation field will not reverse the radial in-fall of matter but {rather} a stalemate between the radiation and gravitational field will be achieved, {where} in-fall is halted but not reversed. Certainly, this picture is not at all congruent (or somewhere near there) to the common picture that is {accepted} in the popular literature where at $\sim8-10\mathcal{M}_{\odot}$, all the circumstellar material{,} from the surface of the {star right} up to the edge of the {molecular cloud} core{,} is expected to be swept away by the powerful radiation field. This finding is not a complete but {rather} a partial solution to the radiation problem in that beyond the $8-10\mathcal{M}_{\odot}$ limit, the nascent star will not accrete any further{. Under this model,} its mass will {stay at} this value{; it} accret{es} from the stagnant and frozen envelope once its mass drops below this $8-10\mathcal{M}_{\odot}$ limit. {A v}ery important {point} to note is that this is for a spherically symmetric gravitational setting where the gravitational field {only has} the radial dependence and is {exactly} described by Newton's inverse square law.

In a different reading{,} Nyambuya (\cite{nyambuya10a}){,}  an Azimuthally Symmetric Theory of Gravitation (ASTG) was set-up and {thereby}  a thesis was set-forth to the effect {that:} ($1$) for a non-spinning star, its gravitational field is spherically symmetric, {so}  it  {depends} on the radial distance from the central body; ($2$) for a spinning gravitating body, the gravitational field of the body in question is azimuthally symmetric, {that is to say}, it is dependent on the radial distance from the central body and as-well the azimuthal angle. In a follow-up reading {of} Nyambuya (\cite{nyambuya10b}){,} it has been shown that the ASTG predicts ($1$) that  bipolar outflows may very well be a purely gravitational phenomenon and also that; ($2$) along the spin-equator of a spinning gravitating body, gravity will channel matter onto the spinning nascent star \textit{via} the spin-equatorial disk without radiation having to reverse this inflow, thus allowing stars beyond the critical mass  {$8-10\mathcal{M}_{\odot}$ to} come into {existence}. 

If the ASTG proves itself, then the present reading together with Nyambuya (\cite{nyambuya10a}, \cite{nyambuya10b}) comprise (in our view) a solution to the radiation problem. Given  that the solution to this problem has been sought \textit{via} sophisticated computer simulations and lengthy numerical solutions, and additionally, given  the simplicity and {na\"ivity} of the present approach which seeks to further our understanding of this {problem,  perhaps} this reading presents not only my misunderstanding of the problem, but also of the approach to the problem{. B}ut more on the optimistic side of things, I believe the radiation problem {as discussed herein} has been understood and that the approach is mathematically and physically legitimate, so much that we {hold} the objective view  {that to they} {[i.e. other researchers]} that {seek} a solution to this problem,  {this reading} is something worthwhile.

\section[\textbf{The Radiation Problem}]{The Radiation Problem\label{rp}}

Following Yorke  \cite{yorke02}; for direct radial accretion and accretion \textit{via} the disk to occur onto the nascent star,  {it} is required that the Newtonian gravitational force, $G\mathcal{M}_{star}(t)/r^{2}$, at a point distance $r$ from the star of mass $\mathcal{M}_{star}$ and  luminosity $\mathcal{L}_{star}(t)$ at any time $t$, must {exceed} the radiation force $\kappa_{eff} \mathcal{L}_{star}(t)/4\pi cr^{2}$ \textit{i.e.}: 

\begin{equation}
\frac{G\mathcal{M}_{star}(t)}{r^{2}}>\frac{\kappa_{eff} \mathcal{L}_{star}(t)}{4\pi cr^{2}},\label{rcondition}
\end{equation}

\noindent where  $c=2.99792458\times10^{8}\, \textrm{ms}^{-1}$ is the speed of light in {a} vacuum, $G=6.667\times10^{-11}\,\textrm{kg}^{-1}\textrm{m}^{3}\textrm{s}^{-2}$ is Newton's universal constant of gravitation, $\kappa_{eff}$ is the effective opacity which is the measure of the {gas'} state of being opaque or a measure of the {gas'} imperviousness {to light rays} and is measured in $\textrm{m}^{2}\textrm{kg}^{-1}$. This analysis by Yorke (\cite{yorke02}){,} which is also  reproduced in Zinnecker \textit{\&} Yorke (\cite{zinnecker07}),  is a standard and well accepted analysis that assumes spherical symmetry and{,} at the {same time,} it does not take into account the nascent star's circumstellar material. On the other hand, star formation is not a truly {spherically symmetric} phenomenon (see \textit{e.g.} reviews by Zinnecker \textit{\&} Yorke \cite{zinnecker07};  McKee \textit{\&} Ostrikker \cite{mckee07}) but this simple calculation suffices in as far as defining {the} curtain-region of $8-10\mathcal{M}_{\odot}$ when radiation pressure is expected to become a significant player {in} the {star's formation}. What will be done in this reading is {simply} to perform the same calculation albeit with the circumstellar material taken into account. In the penultimate of this section, we shall make our case based on the {above statements}.

Now, this calculation by Yorke (\cite{yorke02}) and Zinnecker \textit{\&} Yorke (\cite{zinnecker07}), proceeds as follows{:} the inequality (\ref{rcondition}), sets a maximum condition for accretion of material, namely $\kappa_{eff}<4\pi cG\mathcal{M}_{star}(t)/\mathcal{L}_{star}(t)$, and evaluating this we {obtain}: 

\begin{equation}
\kappa_{eff} < 1.30\times10^{4}\left(\frac{\mathcal{M}_{star}(t)}{\mathcal{M}_{\odot}}\right)\left(\frac{\mathcal{L}_{star}(t)}{\mathcal{L}_{\odot}}\right)^{-1},
\end{equation}

\noindent where $\mathcal{M}_{star}(t)$ and $\mathcal{L}_{star}(t)$ are in solar units. Given that, $\mathcal{L}_{star}(t)=\mathcal{L}_{\odot}\left(\mathcal{M}_{star}(t)/\mathcal{M}_{\odot}\right)^{3}$, this implies that: 

\begin{equation}
\kappa_{eff} < 1.30\times10^{4}\left(\frac{\mathcal{M}_{star}(t)}{\mathcal{M}_{\odot}}\right)^{-2}\Rightarrow \left(\frac{\mathcal{M}_{star}}{\mathcal{M}_{\odot}}\right)>\left(\frac{1.30\times10^{4}}{\kappa_{eff}}\right)^{1/2}.\label{rad-enq}
\end{equation}

Now, given that the dusty Interstellar Medium's (ISM) averaged opacity is measured to be about $20.0$ $\textrm{m}^{2}\textrm{kg}^{-1}$ (Yorke \cite{yorke02}) and using this (as an estimate to setting the minimum critical mass, see Yorke \cite{yorke02}; Zinnecker \textit{\&} Yorke \cite{zinnecker07}), we find that this sets a minimum upper mass limit for stars of about $10\mathcal{M}_{\odot}$ for gravitation to dominate the scene before radiation does. It is clear here that the opacity of the molecular cloud material is what sets the critical mass, thus  a cloud of lower opacity will have a higher critical mass. It is expected that the opacity  inside the cloud will be lower than in {the} ISM{.} {I}n adopting the value $\kappa_{eff}=20.0\,\textrm{m}^{2}\textrm{kg}^{-1}$ (see Yorke \cite{yorke02}; Zinnecker \textit{\&} Yorke \cite{zinnecker07}), this was done only to set a minimum lower bound for massive stars. Dust and gas opacities are significantly frequency-dependent and one has to take  this into account for a more rigid {constraint} of a minimum mass for when the radiation field is expected to overcome the gravitational field.

As can be found in Yorke (\cite{yorke02}), the AAM finds some of its {ground} around the alteration of the opacity. For example, if the opacity inside the gas cloud is significantly lower than the ISM value, then accretion can proceed \textit{via} the AAM. To reduce the opacity inside the gas \textit{\&} dust cloud, the AAM posits as one of the its options that optical and Ultra-Violet (UV) radiation inside the accreting material is shifted from the optical/UV into the far Infrared (IR) and also that the opacity may be lower than the ISM value because the opacity will be reduced by the accretion of optically thick material in the blobs of the accretion disk. {Thus reducing} the opacity{,} or finding a physical mechanism that reduces the opacity to values lower than the ISM{,} is a viable solution to the radiation problem. The above mechanism to reduce the opacity {is} rather {ad hoc} and dependent on the environment. 

Now that we have presented the radiation problem as it is commonly understood, we are ready to make our case by inspecting (\ref{rcondition}). Clearly and without any doubt, the left hand side of this inequality is the gravitational field intensity for a gravitating body in empty space while the right hand side is the radiation field of this same star in empty space. From this{,} clearly{,} we are actually comparing the radiation and gravitational field intensity of a star in empty space, whereas {for} the real setting in \textit{Nature}, stars are {found  heavily} enshrouded by gas and dust. Clearly, the conclusions that one finds from (\ref{rcondition}) such as that{,} at about $8-10\mathcal{M}_{\odot}${,} the radiation field of the nascent star is powerful enough to not only halt but reverse the in-fall of material onto the nascent star; this {cannot} be extended to the scenario where a star is submerged in gas and dust{. I}t is erroneous to do so. Clearly, at this very simplistic, {naive} and fundamental level, there is a need to redefine the radiation problem by  including in the left hand side of (\ref{rcondition}), the circumstellar material. Wolfire \textit{\&} Cassinelli (\cite{wolfire87}) {among} others, have performed this calculation where they have taken into account the circumstellar material and {reached} similar conclusions (as \textit{e.g.} those of Yorke \cite{yorke02}). We reach a different conclusion to that of Wolfire \textit{\&} Cassinelli (\cite{wolfire87}) because{,} unlike {these} researchers{,} we use the observational fact that molecular clouds and molecular cores are found exhibiting a well behaved density profile $\rho\propto r^{-\alpha_{\rho}}$, and from this, we {calculate a} general mass distribution ($\mathcal{M}\propto r^{-\alpha}$){. W}e use this to compare the gravitational and radiation field strengths at point $r$ and from there draw our interesting conclusions.

\section[\textbf{Radiation and the Circumstellar Material}]{Radiation and the Circumstellar Material\label{sol}}

Neglecting thermal{,} magnetic {effects}, turbulence and any other forces (as will be shown {latter} in this section, these forces do not change the essence of our argument, hence {we do not} need to worry about them here) and considering only the gravitational and radiation field from the nascent star, we assume here that a star is formed from a gravitationally bound system of material enclosed in a volume space of radius $\mathcal{R}_{core}(t)$ and we shall call this system of material the core and further assume that this core shall have a {constant total} mass $\mathcal{M}_{core}$ at all times. Now{,} as long as the material enclosed in the sphere of radius $r<\mathcal{R}_{core}(t)$ is such that:

\begin{equation}
\frac{G\mathcal{M}(r,t)}{r^{2}}>\frac{\kappa_{eff} \mathcal{L}_{star}(t)}{4\pi cr^{2}},\label{enq1}
\end{equation}

\noindent then, radiation pressure will not exceed the gravitational force in the region $r<\mathcal{R}_{core}(t)${, thus} direct radial in-fall is expected to continue in that region. If $\mathcal{M}_{csl}(r,t)$ is the mass of the circumstellar material at time $t$ enclosed in the region stretching from the surface of the star to the radius $r$, then, $\mathcal{M}(r,t)=\mathcal{M}_{csl}(r,t)+\mathcal{M}_{star}(t)${. Hence,} the difference between (\ref{enq1}) and (\ref{rcondition}) is that in  (\ref{enq1}) we have included the circumstellar material. This is not the whole story.   

Now, (\ref{enq1}) can be written differently as:

\begin{equation}
\mathcal{M}(r,t)>\frac{\kappa_{eff} \mathcal{L}_{star}(t)}{4\pi Gc},\label{enq2}
\end{equation}

\noindent which basically says as long as the amount of matter enclosed in the region of sphere radius $r$ satisfies the above condition, the radiation force will not exceed the gravitational force in that region of radius $r$. In fact, (\ref{enq2}) is the Eddington limit applied to the region of radius $r$. This is identical to equation ($10$) in Wolfire \textit{\&} Cassinelli (\cite{wolfire87}). In their work, Wolfire \textit{\&} Cassinelli (\cite{wolfire87}) solve numerically the radiative transfer problem to determine the effective opacity at the outer edge of the massive star forming core and{,} from this{,} they determine the limits {of} grain-sizes that are needed for the formation of massive stars. Wolfire \textit{\&} Cassinelli (\cite{wolfire87})'s approach is a typical approach used to probe the conditions necessary for massive stars to form.

Our approach is very different from that of Wolfire \textit{\&} Cassinelli (\cite{wolfire87}) and most typical approaches used to study the radiation problem where sophisticated computer simulations and numerical solutions are used. Ours is a simple and na\"{i}ve approach needing no computer simulations nor numerical codes. We shall insert $\mathcal{M}(r,t)=\mathcal{M}_{csl}(r,t)+\mathcal{M}_{star}(t)$ into (\ref{enq1}) and {after}  rearranging, one obtains:

\begin{equation}
\mathcal{M}_{csl}(r,t)>\left[\frac{\kappa_{eff} \mathcal{L}_{star}(t)}{4\pi c G\mathcal{M}_{star}(t)}-1\right]\mathcal{M}_{star}(t)=\left[\left(\frac{\mathcal{M}_{star}(t)}{10\mathcal{M}_{\odot}}\right)^{2}-1\right]\mathcal{M}_{star}(t){.}
\end{equation}

\noindent {O}ur main thrust is to seek values of $r$ in the above inequality that satisfy it. We shall do this by finding a form for $\mathcal{M}_{csl}(r,t)$. 

Before doing this, let us apply (\ref{enq2}) to the entire core{,} {that is} $r=\mathcal{R}_{core}$. This must give us the condition when the star's radiation field is strong enough to sweep away all the circumstellar material from the surface of the star right up to the {outer edge} of the core{. In} doing {so}, one finds that the star's luminosity  should be such that:

\begin{equation}
\mathcal{M}_{core}>\frac{\kappa_{eff} \mathcal{L}_{star}(t)}{4\pi  Gc}.\label{enq3}
\end{equation}

\noindent In making this calculation, we have made the tacit and fundamental assumption that the star's mass will continue to increase until the star reaches a critical luminosity determined by the mass of the core{. L}et us denote this critical luminosity by $\mathcal{L}^{*}_{core}$. From the above,  it follows that:

\begin{equation}
\mathcal{L}^{*}_{core}=\frac{4\pi cG\mathcal{M}_{core}}{\kappa_{eff}}.
\end{equation}

\noindent With this {definition},  then for the radiation field to globally overcome the gravitational field, the nascent star's luminosity must exceed the critical luminosity of the core, \textit{i.e.}:

\begin{equation}
\mathcal{L}_{star}(t)>\mathcal{L}^{*}_{core}.
\end{equation}

Now,  knowing the mass-luminosity relationship of stars is given by $\mathcal{L}_{star}(t)=\mathcal{L}_{\odot}\left(\mathcal{M}(t)/\mathcal{M}_{\odot}\right)^{3}$, then the critical condition $\mathcal{L}_{star}(t)=\mathcal{L}^{*}_{core}$ will occur when:

\begin{equation}
\left(\frac{\mathcal{M}_{star}}{\mathcal{M}_{\odot}}\right)=\left(\frac{\kappa_{eff}\mathcal{L}_{\odot}}{4\pi G\mathcal{M}_{\odot} c}\right)^{-1/3}\left(\frac{\mathcal{M}_{core}}{\mathcal{M}_{\odot}}\right)^{1/3}.
\end{equation}

\noindent Given this and taking $\kappa_{eff}=20.0\,\textrm{m}^{2}\textrm{kg}^{-1}$ and then {plugging these} and the other relevant values{, such as} $G,c$, \textit{etc}{,} into the above, we are lead to:

\begin{equation}
\left(\frac{\mathcal{M}_{max}}{\mathcal{M}_{\odot}}\right)=\left(\frac{\mathcal{M}_{core}}{10\mathcal{M}_{\odot}}\right)^{1/3}.\label{mass-cloud}
\end{equation}

\noindent where we have set $\mathcal{M}_{star}=\mathcal{M}_{max}$. As {we} already said, using $\kappa_{eff}=20.0\,\textrm{m}^{2}\textrm{kg}^{-1}$ gives us the minimum lower bound. What this means is that the mass of the core from which a star is formed may very well be {crucial  in} deciding the final mass of the star because the mass of the core determines the time when global in-fall reversal will occur. 

From this simplistic and rather na\"ive calculation, we can estimate the efficiency of the core:

\begin{equation}
\xi_{core}=\left(\frac{\mathcal{M}_{star}}{\mathcal{M}_{core}}\right)=0.10\left(\frac{\mathcal{M}_{core}}{10\mathcal{M}_{\odot}}\right)^{-2/3}\label{astglarson},
\end{equation}

\noindent thus a $100\mathcal{M}_{\odot}$ core will (according to the above) form a star at an efficiency rate of about $2\%$ and it will produce a star of mass $2\mathcal{M}_{\odot}$. A $10\mathcal{M}_{\odot}$ star will be produced by a core of mass $10^{4}\mathcal{M}_{\odot}$ at an efficiency rate of about $0.1\%$. A $10^{4}\mathcal{M}_{\odot}$ core is basically a fully-fledged molecular cloud. The production of this $10\mathcal{M}_{\odot}$ star  is {based} on the assumption that the rest of the material ({$10^{4}\mathcal{M}_{\odot}-10\mathcal{M}_{\odot}$ = $9.99\times 10^{3}\mathcal{M}_{\odot}$}) will not form stars. In reality, some of the material in this $10^{4}\mathcal{M}_{\odot}$ core  will form many other stars. Further{more}, a $100\mathcal{M}_{\odot}$ star will form in a GMC of mass about $10^{7}\mathcal{M}_{\odot}$.  The above deductions{,} that high mass stars will need to form in clouds of mass $\geq10^{4}\mathcal{M}_{\odot}$, {is in resonance} with the observational fact that massive stars are not found in isolation (\textit{e.g.} Hillenbrand \cite{hillenbrand97}; Clarke et al. \cite{clarke00}) since the other material will form stars.

{R}elationship (\ref{mass-cloud}) is interesting {because of} its similarity to Larson's $1982$ empirical discovery. With a handful of data, Larson (\cite{larson82}) was the first to note that the maximum stellar mass of a given population of stars is related to the total mass of the parent cloud from which the stellar population has been born. That is to say, if $\mathcal{M}_{cl}$ is the mass of {a} molecular cloud and $\mathcal{M}_{max}$ is the maximum stellar mass of the population, then:

\begin{equation}
\mathcal{M}_{max}=\left(\frac{\mathcal{M}_{cl}}{\mathcal{M}_{0}}\right)^{\alpha_{L}}\label{larsons-law}
\end{equation}

\noindent where $\mathcal{M}_{0}=13.2\mathcal{M}_{\odot}$ and $\alpha_{L}=0.430$. This law was obtained from a sample of molecular clouds whose masses are in the range $1.30\leq \log_{10} \left(\mathcal{M}/\mathcal{M}_{\odot}\right)\leq 5.50$. Larson's Law is thought {to be} a result of statistical sampling but we are not persuaded to think that this is the case{;} such a coincidence is{,} in our opinion and understanding, to good to be true. We believe Larson's Law is \textit{Nature}'s subtle message to researchers; it is telling us something about the underlaying dynamics of star formation. This said, could the relationship (\ref{mass-cloud}) be related to Larson's result? The indices of Larson's relation and relationship (\ref{mass-cloud}) have a deviation of about $33\%$ and the constant $\mathcal{M}_{0}$ has a similar deviation of about $33\%$. Could Larson's fitting procedure be ``tuned'' to conform to relationship (\ref{mass-cloud}) and if so, does that mean Larson's relationship finds an explanation from this {behavior}? 

Perhaps the deviation of our relation from that of Larson may well be that our result is derived from an ideal situation where we have considered not the other forces{,} such as the magnetic, thermal forces \textit{etc}, {but} als{o w}e have considered star formation as a spherically symmetric process{,  w}hich it is not{,} and this may also be a source of correction to this result in order to bring it {in}to {agreement with} Larson's result. Let us represent all these other forces by $\vec{\textbf{F}}_{other}$ (\textit{e.g.} magnetic, turbulence, {viscocity} \textit{etc}). Clearly these forces will not aid gravity in its endeavor to squeeze all the material to a single point but {rather} aid the radiation pressure in opposing this. Given this,  {we must write Inequality} (\ref{enq1}) as:

\begin{equation}
\frac{G\mathcal{M}(r,t)}{r^{2}}>\frac{\kappa_{eff} \mathcal{L}_{star}(t)}{4\pi cr^{2}}+\frac{|\vec{\textbf{F}}_{other}|}{m},
\end{equation}

\noindent where $m$ is the average mass of the molecular species of the material constituting the cloud. The above can be written in the form:

\begin{equation}
\mathcal{L}_{star}(t)<\frac{4\pi cG\left(\mathcal{M}(r,t)-r^{2}|\vec{\textbf{F}}_{other}|/m\right)}{\kappa_{eff}},
\end{equation}

\noindent and writing $\mathcal{M}^{\prime}(r,t)=r^{2}|\vec{\textbf{F}}_{other}|/m$, we {will} {have:}

\begin{equation}
\mathcal{L}_{star}(t)<\frac{4\pi cG\left[\mathcal{M}(r,t)-\mathcal{M}^{\prime}(r,t)\right]}{\kappa_{eff}}{.}
\end{equation}

\noindent {F}rom this{,} it is clear that the other forces will act in a manner as to reduce the critical luminosity of the core{. T}hus our result (\ref{mass-cloud}), when compared to  natural reality where these other forces are present{, i}s expected {to show}  that deviation from the real observations must occur. As stated in the opening of this section{,} the inclusion of the magnetic {and} thermal forces{,} \textit{etc}{,} will not change the essence of our argument, hence  the above {argument} justifies why we did not have to worry about these other forces {because} the essence of our result {still} stands. The situation is only critical when these other forces become significant in comparison to the gravitational force.

In the succeeding section, we compute the mass distribution function and {then} show that one arrives at the same result as (\ref{enq2}). Additionally and more importantly, we are able to compute the boundaries  where the radiation field will be strong enough to overcome the gravitational field. {Among} other interesting outcomes, we shall see that the radiation field will create a cavity inside the star forming core and that this cavity grows with time in proportion to the radiation field of the nascent star.

\section[\textbf{Mass Distribution Function}]{Mass Distribution Function \label{al}} 

First{,} we compute the enclosed mass $\mathcal{M}(r,t)$. We know that stellar systems such as molecular clouds and core{s} are found {to exhibit} radial density  profiles given by:

\begin{equation}
\rho(r,t)=\rho_{0}(t)\left(\frac{r_{0}(t)}{r}\right)^{\alpha_{\rho}}\label{cprofile}
\end{equation}

\noindent where $\rho_{0}(t)$ and $r_{0}(t)$ are  time dependent normalization constants and $\alpha_{\rho}$ is the density index. In order to make sense of this density profile (\ref{cprofile}){,} w{e  h}ave to calculate these normalization constants. In its bare form, the power law (\ref{cprofile}) as it stands implies an infinite density at $r = 0$. In general, power laws have this property. Obviously, one has to deal with this. The usual or typical way is to impose a minimum value for $r$, say $r=r_{min}=r_{0}(t)$ and, assign a density there. Here, this minimum radius has been made time dependent for the sole reason that if the cloud is undergoing free fall as in the case in star formation regions, this quantity will respond dynamically to this, {so} it will be time dependent. 

Now, for a radially dependent density profile, the mass distribution is calculated from the integral:

\begin{equation}
\mathcal{M}(r,t)= \int^{r}_{r_{min}} 4\pi r^{2}\rho(r,t)dr.\label{intmass}
\end{equation}

\noindent Inserting the density function (\ref{cprofile}) into the above integral and then evaluating the resultant integral, we are {led} to:

\begin{equation}
\mathcal{M}(r,t)= \left(\frac{4\pi\rho_{0}(t)r^{\alpha_{\rho}}_{min}(t)}{3-\alpha_{\rho}}\right)\left(r^{3-\alpha_{\rho}}-r_{min}^{3-\alpha_{\rho}}(t)\right),\label{mdfn}
\end{equation}

\noindent and this formula does not apply to the case $\alpha_{\rho}=3${. T}his is valid for $0\leq\alpha_{\rho}<3$.  The case $\alpha_{\rho}=3$ is described by  a special MDF which i{s} $ \mathcal{M}(r,t)= \left[4\pi\rho_{0}(t)r^{3}_{min}(t)\right]\ln\left(r/r_{min}(t)\right)$. We shall not consider this case as it will not change the essence of our argument. 

Now, what we shall do here is to constrain $\alpha_{\rho}$ and show that: $0\leq \alpha_{\rho} <3$.  This exercise is being conducted to define the domain {in} which our result has physical significance. First we shall establish that $\alpha_{\rho} <3$ and this we shall do by using the method of proof by contradiction. Let ($r_{2}>r_{1}$). For this setting, we expect that [$\mathcal{M}(r_{2})>\mathcal{M}(r_{1})$] {which} is obviou{s b}ecause as {one} zoom{s} out of the molecular clou{d,} one would expect to have {more matter} in {a} bigger sphere of radius {$r_{2}$ t}han that enclosed in {a}  smaller  sphere of radius $r_{1}${. T}herefore{,} our {condition} is: [$r_{2}>r_{1}\Longrightarrow\mathcal{M}(r_{2})-\mathcal{M}(r_{1})\geq0$]. Using equation (\ref{mdfn}), we have:

\begin{equation}
\mathcal{M}(r_{2})-\mathcal{M}(r_{1})= \left(\frac{4\pi\rho_{0}r^{\alpha_{\rho}}_{min}}{3-\alpha_{\rho}}\right)\left(r^{3-\alpha_{\rho}}_{2}-r_{1}^{3-\alpha_{\rho}}\right)>0,\label{md1}
\end{equation}

\noindent and for ($\alpha_{\rho}>3$) we have ($3-\alpha_{\rho}<0$) so when we divide by the term $(4\pi\rho_{0}r^{\alpha}_{min})/(3-\alpha_{\rho})$ on both sides of the inequality, we must change the sign of the inequality from $>$ to $<$ because $(4\pi\rho_{0}r^{\alpha_{\rho}}_{min})/(3-\alpha_{\rho})$ is a negative number. {In s}o doing, we will have: $r^{3-\alpha_{\rho}}_{2}-r_{1}^{3-\alpha_{\rho}}<0$, and this implies $r^{3-\alpha_{\rho}}_{2}<r_{1}^{3-\alpha_{\rho}}$ and from this {directly} follows the relationship:

\begin{equation}
r_{1}<r_{2}{.}\label{md3}
\end{equation}

\noindent {T}his is a clear contradiction because it violates our initial condition [$r_{2}>r_{1}\Longrightarrow\mathcal{M}(r_{2})>\mathcal{M}(r_{1})$] as this is saying [$r_{2}<r_{1}\Longrightarrow\mathcal{M}(r_{2})>\mathcal{M}(r_{1})$] which is certainly wrong. From a purely mathematical stand-point, we are therefore forced to conclude that $\alpha_{\rho}<3$ if the condition [$r_{2}>r_{1}\Longrightarrow\mathcal{M}(r_{2})>\mathcal{M}(r_{1})$] is to hold -- \textit{QED}.

Now we shall establish that $\alpha_{\rho} \geq0$ and we shall do this using physical arguments. If $3-\alpha_{\rho}>3${, a}s one zooms out of the cloud from the center, the cloud's average material density increases. This scenario is unphysical because gravity is an attractive inverse distance law and thus will always pack more and more material in the center than in the outer regions{.  H}ence{,} the only material configuration that can emerge from this setting is one in which the average density of material decreases as one zooms out of the cloud. This implies $3-\alpha_{\rho}\leq 3$ which leads to $\alpha_{\rho}\geq 0${. Thus} combining the two results, we are going to have: $0\leq\alpha_{\rho}<3$. Now we have defined the physical boundaries of the density profile. 

Now we have to normalize the MDF by imposing some boundary conditions. The usual or traditional boundary condition is to set $\mathcal{M}(r_{min},t)=0$ and this i{n f}act means there will be a cavity of radius $r_{min}(t)$ in the cloud. What we shall do {next} is different from thi{s t}raditional normalization. We shall set $\mathcal{M}(r_{min},t)=\mathcal{M}_{star}$ where $\mathcal{M}_{star}$ is the mass of the central star{,} henc{e}  $r_{min}(t)=\mathcal{R}_{star}(t)$. Thus what we have done is to place the nascent star in the cavity {which means} we must write our MDF as:

\begin{equation}
\mathcal{M}(r,t)= \left(\frac{4\pi\rho_{0}(t)\mathcal{R}_{star}^{\alpha_{\rho}}(t)}{3-\alpha_{\rho}}\right)\left(r^{3-\alpha_{\rho}}-\mathcal{R}_{star}^{3-\alpha_{\rho}}(t)\right)+\mathcal{M}_{star}(t),\label{mdf1}
\end{equation}

\noindent and this applies for $\mathcal{R}_{star}(t)\leq r\leq \mathcal{R}_{core}(t)$.

Now, if the mass enclosed inside the core remains constant throughout, then we must have at $r=\mathcal{R}_{core}(t)$ the boundary condition $\mathcal{M}(\mathcal{R}_{core},t)=\mathcal{M}_{core}$. We know that the sum total of all the circumstellar material at any given time is given by: $\mathcal{M}_{csl}(t)=\mathcal{M}_{core}-\mathcal{M}_{star}(t)$. Combining all the information, we will have:

\begin{equation}
\left(\frac{4\pi\rho_{0}(t)r^{\alpha_{\rho}}_{0}(t)}{3-\alpha_{\rho}}\right)=\frac{\mathcal{M}_{csl}(t)}{\mathcal{R}^{3-\alpha_{\rho}}_{core}(t)-\mathcal{R}_{star}^{3-\alpha_{\rho}}(t)},
\end{equation}

\noindent and this means the MDF can now be written as:

\begin{equation}
\mathcal{M}(r,t)= \overbrace{\mathcal{M}_{csl}(t) \left(\frac{r^{3-\alpha_{\rho}}-\mathcal{R}_{star}^{3-\alpha_{\rho}}(t)}{\mathcal{R}^{3-\alpha_{\rho}}_{core}(t)-\mathcal{R}_{star}^{3-\alpha_{\rho}}(t)}\right)}^{\tiny\textbf{Circumstellar}\,\, \textbf{Material}\,\,\textbf{in}\,\,\textbf{Region}\,\,\textbf{Radius}\,\,\textbf{r}}+\overbrace{\mathcal{M}_{star}(t)}^{\tiny\textbf{Mass} \,\, \textbf{of}\,\,\textbf{the}\,\,\textbf{nascent}\,\,  \textbf{star}}\,\,\,\,\,\,\,\,\,\textrm{for}\,\,\,\,\,\,r\geq\mathcal{R}_{star}(t)\label{mdf}.
\end{equation}

\noindent We shall take this as the final form of our mass distribution function. If the reader accepts this, then what follows is {a} straight forward exercise and leads to what we believe is a significant step forward in the resolution of the radiation problem. The reader may want to query that we have overstretched our boundary limits by making the {assumption that the} MDF be continuous from the surface of the star right up {to} the edge of the core. In that event, we need to {make this point} clea{r a}nd reach an accord. 

First{,} let us consider a serene molecular core {long} before a star begins to form at the {center}. We know that the density is not a fundamental physical quantity but a physical quantity derived from two fundamental physical quantities which are mass and volume, \textit{i.e.}, density=mass/volume{. W}e must note that this is defined for (volume$>0$). We shall assume that this core exhibits the density profile $\rho\propto r^{-\alpha_{\rho}}$. This fact that $\rho\propto r^{-\alpha_{\rho}}${, when} combined with the fact that density {is} not a fundamental physical quantity but {a} quantity derived from two fundamental quantities, suggests tha{t} at any given time the mass must be distributed in proportion to the radius, \textit{i.e.}, $\mathcal{M}(r,t)\propto r^{\alpha}$. The radial dependency of the density is an indicator that that mass has a radial dependency. The relationship $\mathcal{M}(r,t)\propto r^{\alpha}$ means we must have $\mathcal{M}(r,t)= ar^{\alpha}+b$ where ($a,b$) are constants. We expect that $\mathcal{M}(0,t)=0$. If this is to hold (as it must), then ($b=0$) and ($\alpha\geq0$). We also expect the condition $\mathcal{M}(\mathcal{R}_{core},t)=\mathcal{M}_{core}$ to hold. If this is to hold (as it must), then we will have $a=\mathcal{M}_{core}/\mathcal{R}_{core}^{\alpha}(t)$ hence $\mathcal{M}(r,t)= \mathcal{M}_{core}(r/\mathcal{R}_{core}(t))^{\alpha}$. From the definition of density this means:

\begin{equation}
\rho(r,t)=\left(\frac{3\mathcal{M}_{core}}{4\pi \mathcal{R}_{core}^{\alpha}(t)}\right)r^{\alpha-3}\,\,\,\,\,\,\,\,\textrm{for}\,\,\,\,\,\,\,\,\,\,r>0.
\end{equation}

Now, if the density profile is to fall off as $r$ increase{s} as is the case in \textit{Nature}, then ($\alpha-3\leq0$) which implies ($\alpha\leq3$). Combining this with ($\alpha\geq0$) we will have ($0\leq\alpha\leq3$). Comparing this with the profile ($\rho\propto r^{-\alpha_{\rho}}$), we have: ($-\alpha_{\rho}=\alpha-3$) and substituting this into ($0\leq\alpha\leq3$), one obtains ($0\leq3-\alpha_{\rho}\leq3$). From ($3-\alpha_{\rho}\leq3$), we have ($\alpha_{\rho}\geq0$), and from ($0\leq3-\alpha_{\rho}$), we have ($\alpha_{\rho}\leq3$), hence ($0\leq\alpha_{\rho}\leq3$).

Now, in this serene molecular {cloud}, a small lamp begins to form{;} let this lamp have a radius $\mathcal{R}_{lamp}(t)$ and mass $\mathcal{M}_{lamp}$. I shall {pose} a question: do we expect this lamp to cause any fundamental changes to the mass distribution $\mathcal{M}(r,t)= ar^{\alpha}+b$? I think not. If this is the case, then our mass distribution must now be defined up {to} the radius of the lamp{,} $\mathcal{M}(\mathcal{R}_{lamp},t)= \mathcal{M}_{lamp}$ and this condition leads to: $b=\mathcal{M}_{lamp}-a\mathcal{R}_{lamp}^{\alpha}${, t}hus: 

\begin{equation}
\mathcal{M}(r,t)= a(r^{3-\alpha_{\rho}}-\mathcal{R}_{lamp}^{3-\alpha_{\rho}})+\mathcal{M}_{lamp}\,\,\,\,\,\,\,\,\,\textrm{for}\,\,\,\,\,\,r\geq\mathcal{R}_{lamp}(t).
\end{equation}

\noindent where we have substituted $\alpha=3-\alpha_{\rho}$. Now inserting the condition that $\mathcal{M}(\mathcal{R}_{core},t)=\mathcal{M}_{core}$, we will have:

\begin{equation}
a=\left(\frac{\mathcal{M}(r,t)-\mathcal{M}_{lamp}}{\mathcal{R}^{3-\alpha_{\rho}}_{core}(t)-\mathcal{R}_{lamp}^{3-\alpha_{\rho}}}\right)\,\,\,\,\,\,\,\,\,\textrm{for}\,\,\,\,\,\,r\geq\mathcal{R}_{lamp}(t).
\end{equation}

\noindent and putting all this together we will have:

\begin{equation}
\mathcal{M}(r,t)= \mathcal{M}_{csl}(t)\left(\frac{r^{3-\alpha_{\rho}}-\mathcal{R}_{lamp}^{3-\alpha_{\rho}}}{\mathcal{R}^{3-\alpha_{\rho}}_{core}(t)-\mathcal{R}_{lamp}^{3-\alpha_{\rho}}}\right)+\mathcal{M}_{lamp}\,\,\,\,\,\,\,\,\,\textrm{for}\,\,\,\,\,\,r\geq\mathcal{R}_{lamp}(t).\label{lamp}
\end{equation}

\noindent where $\mathcal{M}_{csl}(t)=\mathcal{M}_{core}-\mathcal{M}_{lamp}(t)$. Comparison of the above with (\ref{mdf}) shows that the lamp in the above formula is the star in (\ref{mdf}).

 We are certain {that} the reader will have no problem with (\ref{lamp}) because the lamp does not disrupt the mass distribution since it has no radiation{. H}ence{,} we would expect a continuous distribution of mass right up {to} the surface of the lamp as material will be flowing into the lamp. {However,} this same lamp is a protostar and at somepoint it must switch on to become a star. At this moment, assuming the correctness of the thesis that at $8-10\mathcal{M}_{\odot}$, the radiation field begins to push material away from the nascent star, we could from logic expect that the mass distribution must be continuous up till that time {when} disruption starts at $8-10\mathcal{M}_{\odot}$. During the time when the lamp's (or protostar's) mass is in the range $0\leq\mathcal{M}_{lamp}(t)< 8-10\mathcal{M}_{\odot}$, the MDF (\ref{lamp}) must hold. From this, we have just justified the formula (\ref{mdf}) for the mass range: $0\leq\mathcal{M}_{star}(t)< 8-10\mathcal{M}_{\odot}$. When the radiation field begins to be significant, we shall have to check and revise this formula.
 
Now, {from} the MDF (\ref{mdf}), the gravitational field intensity{,} at any given time $t$ and at any given point $r$ inside the core from the surface of the star{,} will be given by: 

\begin{equation}
\vec{\textbf{g}}(r,t) = \overbrace{-\left(\frac{G\mathcal{M}_{csl}(t)}{r^{2}}\right)\left(\frac{r^{3-\alpha_{\rho}}-\mathcal{R}_{star}^{3-\alpha_{\rho}}(t)}{\mathcal{R}^{3-\alpha_{\rho}}_{core}(t)-\mathcal{R}_{star}^{3-\alpha_{\rho}}(t)}\right)\hat{\textbf{r}}}^{\tiny \textbf{Circumstellar}\,\, \textbf{Gravitation}}-\overbrace{\left(\frac{G\mathcal{M}_{star}(t)}{r^{2}}\right)\hat{\textbf{r}}}^{\tiny \textbf{ Star's}\,\, \textbf{Gravitation}}.\label{nng}
\end{equation}

\noindent Clearly,  we have been able to separate the gravitation due to the star {from that due to} the circumstellar material. 

Now, from the above, the inequality (\ref{enq1}) becomes:

\begin{equation}
\left(\frac{G\mathcal{M}_{csl}(t)}{r^{2}}\right)\left(\frac{r^{3-\alpha_{\rho}}-\mathcal{R}_{star}^{3-\alpha_{\rho}}(t)}{\mathcal{R}^{3-\alpha_{\rho}}_{core}(t)-\mathcal{R}_{star}^{3-\alpha_{\rho}}(t)}\right)+\left(\frac{G\mathcal{M}_{star}(t)}{r^{2}}\right)>\frac{\kappa_{eff} \mathcal{L}_{star}(t)}{4\pi r^{2}c},\label{eq}
\end{equation}

\noindent where the first term on the left hand-side of (\ref{eq}) is clearly the gravitational field intensity of the circumstellar material and the second term is the gravitational field of the nascent star.

\section[\textbf{Radiation Cavity}]{Radiation Cavity}

The inequality (\ref{enq2}) gives us {a}  condition that must be met before the radiation field is powerful enough {that} it can push away (all) the circumstellar material inside the shell of radius $r$. Beyond this radius, the radiation field is not at all powerful enough to overcome the gravitational field. Unfortunately, one {cannot} deduce this radius $r$ from (\ref{enq2}). {T}he inequality (\ref{eq}){,} as does (\ref{enq2}) {and} (\ref{eq}) tells us the conditions to be met before the radiation field is powerful enough to halt in-fall{. I}n addition to this, (\ref{eq}) {yields} more information than (\ref{enq2}) because in (\ref{eq}) we have quantified the MDF for the circumstellar material and this allows us to compute the region $r$ where the radiation field is much stronger than the gravitational field.  From (\ref{eq}){,} we deduce that the radiation field will create a cavity in the star forming core; in this cavity, the radiation field is much stronger than the gravitational field{,} thus there will be {a radiation cavity with} no material but {only} radiation{,} hence the term {``radiation cavity''}.  To se{e t}hat (\ref{eq}) {describes} a cavity, we simpl{y} have to write (\ref{eq}) with $r$ as the subject of the formula; {after} doing {so}{,} one arrives at:

\begin{equation}
r>\left(\frac{\left(\kappa_{eff} \mathcal{L}_{star}(t)-{4}\pi cG\mathcal{M}_{star}(t)\right)\left(\mathcal{R}^{3-\alpha_{\rho}}_{core}(t)-\mathcal{R}_{star}^{3-\alpha_{\rho}}(t)\right)}{{4}\pi c G\mathcal{M}_{csl}(t)}+\mathcal{R}_{star}^{3-\alpha_{\rho}}(t)\right)^{\frac{1}{3-\alpha_{\rho}}}=\mathcal{R}_{cav}(t) \label{eq1}
\end{equation}

\begin{figure}
\centering
\includegraphics[scale=0.5]{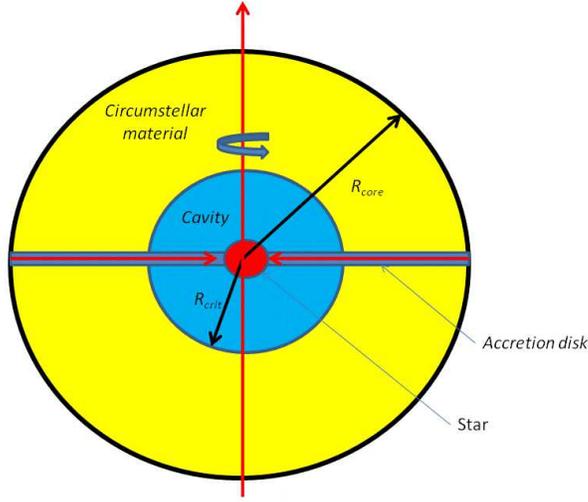}
\caption[An illustration of the Cavity inside the Core, and an Accretion Disk]{For a non-spinning core at $\sim8-10\mathcal{M}_{\odot}$, the nascent stars's accretion is halted (and importantly{,} in-fall is not reversed but only halted) because when the radiation field tries to create a cavity in which process  the star is separated from its accretion source which is the circumstellar material{. T}his means the star's mass accretion is halted because its mass can no longer grow since there exists no other channel(s) \textit{via} which its mass feeds. Should the star's mass fall below $\sim8-10\mathcal{M}_{\odot}$, the circumstellar material will fall onto the nascent star until its mass is restored to its previous value of $\sim8-10\mathcal{M}_{\odot}$. In order for the radiation field to start pushing the circumstellar material, its mass must exceed $\sim8-10\mathcal{M}_{\odot}$. Since there is no way to do this, in-fall is only halted and not reversed.  Henc{e,} the star's mass for a non-spinning star stays {constant} at $\sim8-10\mathcal{M}_{\odot}$. As urged in Nyambuya (\cite{nyambuya10b}), this scenario is different for a spinning star because the ASGF (which comes about due the spin of the nascent star) allows  matter to continue  accreting \textit{via} the equatorial disk inside the cavity as illustrated above. The accretion disk will exist inside the radiation cavity and{, a}ccording to the azimuthally symmetric theory of gravitation (Nyambuya \cite{nyambuya10b}), {this disk should} channel mas{s o}nto the nascent star right-up to the surface of the star without radiation hindrances.}
\label{radfig}
\end{figure}

\noindent where $\mathcal{R}_{cav}(t) $ is the radius of the cavity. Now that there is a cavity{, let us} pause so that we can revise the MDF. Clearly, in the case where there are outflows, this must be given by:

\begin{equation}
\mathcal{M}(r,t)= \mathcal{M}_{csl}(t)\left(\frac{r^{3-\alpha_{\rho}}-\mathcal{R}_{cav}^{3-\alpha_{\rho}}}{\mathcal{R}^{3-\alpha_{\rho}}_{core}(t)-\mathcal{R}_{cav}^{3-\alpha_{\rho}}}\right)+\mathcal{M}_{*}(t)\,\,\,\,\,\,\,\,\,\textrm{for}\,\,\,\,\,\,r\geq\mathcal{R}_{cav}(t).\label{cavmdf}
\end{equation}

\noindent where $\mathcal{M}_{csl}(t)=\mathcal{M}_{core}-\mathcal{M}_{*}(t)$ and $\mathcal{M}_{*}(t)=\mathcal{M}_{star}(t)+\mathcal{M}_{disk}(t)+\mathcal{M}_{outf}(t)$: $\mathcal{M}_{disk}(t)$ is the disk mass inside the cavity at time $t$ and $\mathcal{M}_{outf}$ the bipolar outflow contained in the cavity at time $t$.

Now, what this inequality (\ref{eq1})  is ``\textit{saying}'' is that, at any given moment {when} the star has surpassed the critical mass ($8-10\mathcal{M}_{\odot}$), there will exist a region $r< \mathcal{R}_{cav}(t)$ where the radiation field will reverse the radially in-falling material and in the region $r> \mathcal{R}_{cav}(t)$, for material therein, the radiation field has not reached a state where it exceeds the gravitational field{. H}ence{,} in-fall reversal in that region has not been achieved. This region [\textit{i.e.} $r< \mathcal{R}_{cav}(t)$] grows with time thus the radiation field slowly and gradually pushe{s t}he material further and further away from the nascent star until $\mathcal{R}_{cav}(t)=\mathcal{R}_{cl}$ where radial in-fall is completely halted{. T}his will occur when the star has reached the critical core luminosity $\mathcal{L}^{*}_{core}$. The condition when the critical core luminosity has been shown earlier  lead{s} to (\ref{astglarson}) which is a Larson-like relation{, \textit{i.e.}} (\ref{larsons-law}), \textit{ipso facto}{, t}his strongly suggests that Larson's Law may not be a result of statistical sampling but a statement {(about)} an{d a} fossil record of the battle of forces between gravitation and the radiation field.

By saying that the nascent massive star will create a cavity, we have made a tacit and fundamental assumption that its mass will continue to grow soon after  the cavity begins to form and that its mass will thereafter continue to grow while in the cavity. {However,} how can this be since the cavity separates the nascent star from the circumstellar matter? The nascent star { now does not have} a channel to feed its mass{,} so there can be no growth in its mass unless there exists a channel \textit{via} which its mass feeds. At this juncture, we direct the reader to the readings Nyambuya (\cite{nyambuya10b}, \cite{nyambuya10a}).

In Nyambuya (\cite{nyambuya10a}), as already said in the introductory section, we set-up the ASTG {such that} the thesis was advanced to the effect ($1$) that,  for a non-spinning star, its gravitational field is spherically symmetric (to be specific, i{t o}nly depend{s} on the radial distance from the central body); ($2$) that, for a spinning gravitating body, the gravitational field of {the} body in question is azimuthally symmetric, \textit{i.e.}, i{t d}epend{s} on the radial distance ($r$) from the central body an{d t}he azimuthal angle ($\theta$). In a follow-up reading{,} Nyambuya ($2010b$); we showed that the ASTG predicts ($1$) that  bipolar outflows may very well be a purely gravitational phenomenon (\textit{i.e.}, a repulsive gravitational phenomenon) and also that; ($2$) along the spin-equator (define{d} Nyambuya \cite{nyambuya10b})  of a spinning gravitating body, gravity will channel matter onto the spinning nascent star \textit{via} the accretion disk (l{y}ing along the spin-equator) thus allowing stars beyond the critical mass  $8-10\mathcal{M}_{\odot}$ {t}o {form and begin their stellar processes}. It should be said tha{t a}ccretion {disks} can also be formed by a number of different mechanisms other than {an} Azimuthally Symmetric Gravitational Field (ASGF).

The accretion of matter beyond the $8-10\mathcal{M}_{\odot}$ limit must only be possible for a spinning star because it possesses the ASGF that is needed to continue the channeling of matter onto the star \textit{via} the accretion disk  -- see {the} illustration in figure \ref{radfig}. For a non-spinning core{,} the nascent stars's accretion {cannot} proceed beyond $8-10\mathcal{M}_{\odot}${. I}t is halted because the moment the radiation field tries to create a cavity{,} when the (non-spinning) star's mass is $8-10\mathcal{M}_{\odot}$, the (non-spinning) sta{r  t}hat very moment becomes separated from the surrounding circumstellar material. This means the (non-spinning) star's mass accretion is halted because its mass can no longer grow since there exists no other channel(s) \textit{via} which its mass feeds. Should the (non-spinning) star's mass fall below $8-10\mathcal{M}_{\odot}$, the circumstellar material will fall onto the nascent (non-spinning) star until its mass is restored to its previous value of $8-10\mathcal{M}_{\odot}$. This means the star's mass for a non-spinning star stays {constant} at $8-10\mathcal{M}_{\odot}$. As explained in the above paragraphs, this scenario is different for a spinning star because the ASGF (which comes about due {to} the spin of the nascent star) allows  matter to continue  accreting \textit{via} the equatorial disk. The accretion disk will exist inside the radiation cavity and this disk should according to the ASTG (Nyambuya \cite{nyambuya10b}){,} channel mass right up to the surface of the star without radiation hindrances. The scenario just present{ed} is completely different from that projected in much of the wider literature where{,} at $8-10\,\mathcal{M}_{\odot}$, suddenly the radiation is so powerful {that} it reverses any further in-fall.  It is bona-fide knowledge that star formation is not a spherically symmetric process and from the above, it follows that stars beyond the $8-10\mathcal{M}_{\odot}$ limit must {form} with no hindrance{,} form the radiation field and the only limit to their existence is {if the} gravitationally bound core {has} enough mass to form them. 

\section[\textbf{Discussion and Conclusions}]{Discussion \textit{\&} Conclusions}

This contribution coupled with  Nyambuya (\cite{nyambuya10b}) {seem} to strongly point to the possibility that the radiation problem of massive stars may not exist as previously thought. In the present reading, we find that beginning at the time when $\mathcal{M}_{star}(t)\simeq8-10\mathcal{M}_{\odot}$, the radiation field will create a cavity  inside the star forming core and the circumstellar material inside the region $\mathcal{R}_{cav}(t)<r\leq\mathcal{R}_{core}(t)$ is going to be pushed  gradually ({in particular}, not blown away) as the radiation field from the star grows until a point is reached when the cavity is the size of the core itself{. At this} point{,} complete in-fall reversal is attained. If the radiation field of the star is to grow, its mass must grow, thus, the cavity must not prevent accretion of mass onto the nascent star and this is possible for a spinning massive star. Once the cavity is created, the mass of the nascent will{,} for a spinning massive star; feed \textit{via} the accretion disk and this disk is not affected by the radiation field. By saying the disk is not affected by the radiation field{,} we mean the material on the disk is not going to be pushed away by the radiation field as it pushes the other material away because the azimuthally symmetric gravitational field of the star is powerful enough along this plane to overcome the radiation field{. T}his has been shown or argued in  Nyambuya (\cite{nyambuya10b}) that this must be the case.

The ASGF is only possible for a spinning star; since all known stars are spinning, every star should{,} according to the ASTG{,} have the potential to grow to higher masses. This mean{s m}assive stars should {start their stellar processes} because of their spin which bring{s} about the much needed ASGF. A {non-spinning} {star will} have no ASGF, hence {there will be} no disk around it {to} {channel material} once the radiation field begins {taking} its toll. In this case of a {non-spinning} star{,} once the star has reached the critical mass $\sim8-10\mathcal{M}_{\odot}$, its mass {cannot} grow any further because {at} the moment it tries to grow, the star and the circumstellar material become separated due to the radiation field which{,} in this case{,} is stronger tha{n} the gravitational field. In this event, any further growth in mass of the star is {stymied}. This{,} i{n  f}act{,} means tha{t a}s long as there is circumstellar material, the mass of a {non-spinning} star will stay {constant} at $\sim8-10\mathcal{M}_{\odot}$ because{,} the moment it fall{s  s}lightly below $\sim8-10\mathcal{M}_{\odot}$, gravity becomes more powerful{,} thus accret{ing only enough mass} to restore it to its previous value of $\sim8-10\mathcal{M}_{\odot}$. In this case, we have an ``eternal'' stalemate between the gravitational and radiation field.

An important and subtle difference between the present work and that of other researchers (Larson \textit{\&} Starfield \cite{larson71}; Kahn \cite{kahn74}; Yorke \textit{\&} Kr$\ddot{\textrm{u}}$gel \cite{yorke74};  Wolfire \textit{\&} Cassinelli \cite{wolfire87};  Palla \textit{\&} Stahler \cite{palla93}; Yorke \cite{yorke02}; Yorke \textit{\&} Sonnhalter \cite{yorke-s03}) is that we have seized on the observational fact that molecular clouds and cores are found  exhibiting well defined density profiles. From this we computed the MDF which enabled us to  {exactly} find the physical boundaries where the gravitational field is expected to be much stronger than the radiation field once the star exceed the critical mass. Additionally and more importantly is that from Nyambuya (\cite{nyambuya10b}) we have been able to argue that even after the cavity has been created mass will be channeled on to the star \textit{via} the accretion disk. Without the ideas presented in Nyambuya (\cite{nyambuya10b}), we would have been stuck because we where going to find oursel{ves} without a means to justify how the mass accretion continues once the cavity has been created. 

Importantly, we have pointed out a real problem in Yorke (\cite{yorke02}), Yorke \textit{\&} Sonnhalter (\cite{yorke-s03}) and Zinnecker \textit{\&} Yorke (\cite{zinnecker07}), namely that  these researchers have neglected the treatment of the circumstellar material in their theoretical arguments leading to their definition of the radiation problem because they used Newton's inverse square law which clearly applies to a non-rotating mass in empty space{, so} the inequality (\ref{enq1}) applies only for a star in empty space. In empty space, it is correct to say that the radiation field for a star of  mass $10\,\mathcal{M}_{\odot}$ and beyond, will exceed the gravitational field everywhere in space beyond the nascent star's surface, but the same is not true for a star submerged in a pool of gas{, which} as is the case {for} the stars  that we observe.

Another important outcome is that it appear{s  L}arson's Laws may well be a signature and fossil record of the battle of forces between the radiation and gravitational field{s}. At present, it is thought of as being a result of statistical sampling{. T}hus the present brings us to start rethinking this view. We are not persuaded to think this is a result of statistical sampling. This view finds support from Weidner et al. (\cite{weidner09})'s most recent and exciting work. In this work, these researchers present a thorough literature study of the most-massive star{s} in several young star clusters in order to assess whether or not star clusters are populated from the stellar initial mass function (IMF) by random sampling over the mass range ($0.01\mathcal{M}_{\odot}\leq \mathcal{M}_{star}\leq 150\mathcal{M}_{\odot}$) without being constrained by the cluster mass. Their  data reveal a partition of the sample into lowest mass objects ($\mathcal{M}_{cl}\leq 100\mathcal{M}_{\odot}$), moderate mass clusters ($100\mathcal{M}_{\odot}\leq \mathcal{M}_{cl}\leq 1000\mathcal{M}_{\odot}$) and rich clusters above ($ \mathcal{M}_{cl}\geq 1000\mathcal{M}_{\odot}$) where $\mathcal{M}_{cl}$ is the mass of the molecular cloud. Their statistical tests of this data set reveal that the hypothesis of random sampling is highly unlikely{,} thus strongly suggesting that there exists some well defined physical cause. 

In closing,  allow us to say tha{t w}e do not claim to have solved the radiation problem but merely believe that what we have presented herein{, t}ogether with the readings Nyambuya (\cite{nyambuya10b}, \cite{nyambuya10a}){,} is work that  may very well be a significant step forward in the endeavor to resolv{e} this massive star formation riddle.\\

\noindent \textbf{\underline{Acknowledgments}:} I am grateful to my brother, Baba \textit{va} Panashe, and his wife, Amai \textit{va} Panashe, for their kind hospitality they offered while working on this reading and to 
 Mr. Isak D. Davids \textit{\&} Ms. M. Christina Eddington for proof reading the grammar, spelling \textit{\&} language editing. Last and certainly not least, I am very grateful to my Professor, D. Johan van der Walt, and Professor Pienaar Kobus, for the strength and courage that they have given me.

\label{lastpage}

\begin{thebibliography}{100}

\bibitem[$1998$]{bonnell98}
Bonnell I. A., Bate M., Zinnecker H., $1998$, MNRAS, $298$, $93$

\bibitem[${2001}$]{bonnell01}	
Bonnell I. A., Clarke C. J., Bate M. R., Pringle J. E., $2001$, MNRAS, $324$, $573$

\bibitem[${2002}$]{bonnell02}	
Bonnell I. A., Bate M. R., $2002$, MNRAS, $336$, $659$

\bibitem[${2004}$]{bonnell04}
Bonnell I. A., Vine S. G., Bate, M. R., $2004$, MNRAS, $349$, $735$  

\bibitem[$2006$]{bonnell06}
Bonnell I. A., Clarke C. J., Bate M. R., $2006$, MNRAS, $368$, $1296$

\bibitem[$2007$]{bonnell07}
{Bonnell, I. A., Larson, R. B., \&
Zinnecker, H. 2007, Protostars and Planets V, ed. V. B. Reipurth, D. Jewitt, \& K. Keil (Tucson, AZ: Univ.
Arizona Press), 149 (arXiv:0603447)}

\bibitem[$2000$]{clarke00}
Clarke C. J., Bonnell I. A., Hillenbrand L. A., $2000$, Protostars and Planets IV, $151$


\bibitem[$1997$]{hillenbrand97}
Hillenbrand L. A., $1997$, AJ, $113$, $1733$

\bibitem[${2009}$]{krumholz05}	
Krumholz M. R., Klein R. I., McKee C. F. {et al., $2005$}, Science, $323$, ${754}$

\bibitem[${2005}$]{krumholz09}	
Krumholz M. R., Klein R. I., McKee C. F., Offner S. S. R., Cunningham A. J., ${2005}$, Protostars and Planets V, $1286$, $8271$

\bibitem[$1974$]{kahn74}
Kahn F. D., $1974$, A\textit{\&}A, $37$, $149-162$

\bibitem[${1972}$]{larson72}
Larson R. B., $1972$, MNRAS, 156,$437$

\bibitem[$1982$]{larson82}
Larson R. B., $1982$, MNRAS, $200$, $159$

\bibitem[$1971$]{larson71}
Larson R. B. \textit{\&} Star{r}field S., $1971$, {A\&A}, $13$, $190$


\bibitem[${2002}$]{maeder02} 
Maeder A., Behrend R., $2002$, ASP Conf. Ser., $267$, $179$ 

\bibitem[$2007$]{mathews07} 
Matthews L. D., Goddi C., Greenhill L. J., Chandler C. J., Reid M. J. \textit{\&}  Humphreys E. M. L., $2007$, {Astrophysical Masers and their Environments (IAU Symp. $242$), ed. J. M. Chapman \textit{\&} W. A. Baan (Dordrecht: Kluwer), $130$}

\bibitem[$2007$]{mckee07}
McKee C. F., Ostriker E. C., $2007$, {ARA\textit{\&}A}, $45$, $565$ 

\bibitem[${2010a}$]{nyambuya10a}
Nyambuya G. G., ${2010a}$, {MNRAS, 403, Issue 3, 1381.}

\bibitem[${2010b}$]{nyambuya10b}
Nyambuya G. G., ${2010b}$, RAA (Research in Astronomy and Astrophysics), Vol. $10$ No. $11$, $1151-1176$  ($viXra:0911.0025$).

\bibitem[$1993$]{palla93}
Palla F., Stahler S. W., $1993$, AJ, $418$, $414$

\bibitem[$1976$]{shu76}
Shu F. H., $1977$,  Bulletin of the Amer. Astron. Soc., $8$, $547$

\bibitem[$1977$]{shu77}
Shu F. H., $1977$,  AJ, $214$, $488-497$

\bibitem[$2002$]{yorke02} 
Yorke H. W., $2002$, ASP Conf. Series, $267$, $165$ 

\bibitem[${2004}$]{yorke03}
{Yorke, H. W. $2004$, Star Formation at High Angular Resolution (IAU Symp. $221$), ed. M. Burton, R. Jayawardhana, \textit{\&} T. Bourke (Dordrecht: Kluwer), $141$}

\bibitem[${1977}$]{yorke74}
Yorke H. W. \textit{\&} Kr${\ddot{\rm{u}}}$gel, {$1977$, A\textit{\&}A,} $54$,  $183$

\bibitem[${2003}$]{yorke-s03}
Yorke H. W., Sonnhalter, {$2002$}, ApJ, $569$,  $846-862$

\bibitem[$2009$]{weidner09}
Weidner C., Kroupa P., Bonnell I. A. D., $2009$, {MNRAS, 401, 275.}

\bibitem[$1987$]{wolfire87}
Wolfire M. G., Cassinelli J. P., $1987$, ApJ, $319$, $850$

\bibitem[${2007}$]{zinnecker07}
Zinnecker H. \& Yorke H. W., $2007$, {ARA\textit{\&}A}, $45$, $481$

\end{thebibliography}
\end{document}